\pdfoutput=1 
\documentclass[journal]{IEEEtran}

\usepackage{amsmath}
\usepackage{algpseudocode}
\usepackage[pdftex]{graphicx}
\usepackage{subfig}
\usepackage{tikz}
\usetikzlibrary{shapes.geometric, arrows}
\usepackage{amsfonts}
\tikzstyle{startstop} = [circle, rounded corners, minimum width=0.5cm,text centered, draw=black]
\tikzstyle{startstop2} = [circle, rounded corners, minimum width=0.5cm,text centered, draw=black,fill=red!8]
\tikzstyle{startstop3} = [circle, rounded corners, minimum width=0.5cm,text centered, dashed, draw=black,fill=red!0]
\tikzstyle{arrow} = [thick,->,>=stealth]
\tikzstyle{arrow1} = [dashed,->,>=stealth]

\usepackage{amsthm}
\newtheorem{theorem}{Theorem}

\def\y{{\mathbf y}}
\def\h{{\mathbf h}}
\def\H{{\mathbf H}}
\def\e{{\mathbf e}}
\def\s{{\mathbf s}}
\def\v{{\mathbf v}}
\def\I{{\mathbf I}}
\def\R{{\mathbf R}}

\def\V{{\mathbf V}}
\def\W{{\mathbf W}}

\def\Q{{\mathbf Q}}
\def\q{{\mathbf q}}
\def\J{{\mathbf J}}
\def\A{{\mathbf A}}
\def\B{{\mathbf B}}

\def\K{{\mathbf K}}
\def\G{{\mathbf G}}
\def\g{{\mathbf g}}
\def\cH{{\boldsymbol {\cal H}}}
\def\cV{{\boldsymbol {\cal V}}}

\def\cJ{ {\cal J}}

\begin{document}
\title{Structure-Based Subspace Method for Multi-Channel Blind System Identification}

\author{Qadri~Mayyala,~\IEEEmembership{Student Member,~IEEE,}
        Karim~Abed-Meraim,~\IEEEmembership{Senior Member,~IEEE,}
        and~Azzedine~Zerguine,~\IEEEmembership{Senior Member,~IEEE}
\thanks{Q. Mayyala and A. Zerguine are with the Department
of Electrical and Electronic Engineering, King Fahd University of Petroleum \& Minerals, Saudi Arabia, e-mails: \{qmayyala, azzedine\}@kfupm.edu.sa.}
\thanks{K. Abed-Meraim is with the PRISME lab, University of Orl\'eans, France, e-mail: karim.abed-meraim@univ-orleans.fr}

\thanks{The authors acknowledge the support provided by the Deanship of Scientiﬁc Research at KFUPM under Research Grant RG1414.}

\thanks{}} 

\markboth{Submitted to IEEE SIGNAL PROCESSING LETTERS, January 2017}
{Shell \MakeLowercase{\textit{et al.}}: Bare Demo of IEEEtran.cls for IEEE Journals}

\maketitle

\begin{abstract}
In this work, a novel subspace-based method for blind identification of multichannel finite impulse response (FIR) systems is presented. Here, we exploit directly the impeded Toeplitz channel structure in the signal linear model to build a quadratic form whose minimization leads to the desired channel estimation up to a scalar factor. This method can be extended to estimate any predefined linear structure,  e.g. Hankel, that is usually encountered in linear systems. Simulation findings are provided to highlight the appealing advantages of the new structure-based subspace (SSS) method over the standard subspace (SS) method in certain adverse identification scenarii.
\end{abstract}

\IEEEpeerreviewmaketitle

\section{Introduction}

\IEEEPARstart{B}{lind} system identification (BSI) is one of the fundamental signal processing problems that was initiated more than three decades ago. BSI refers to the process of retrieving the channel's impulse response based on the output sequence only. As it has so different applications, such as mobile communication, seismic exploration, image restoration and other medical applications, it has drawn researchers' attention and resulted in a plethora of methods. Since then, a class of subspace-based methods dedicated to BSI has been developed, including the standard subspace method (SS) \cite{moulines1995subspace, kang2005subspace}, the cross-relation (CR) method \cite{xu1995least} and the two-step maximum likelihood (TSML) method \cite{hua1996fast}. According to the comparative studies which have been done early in \cite{qiu1996performance3} and \cite{qiu1996performance}, the SS method is claimed to be the most powerful one.
\par
In this paper, we introduce another subspace-based method based on the channel's Toeplitz structure which is employed directly to formulate our cost function. The Toeplitz structure is an inherent nature that exists in most of the linear systems due to their convolutive nature. 
\par
The paper presentation focuses at first on the development of the proposed structure-subspace (SSS) method. Then, we highlight the improvement that is obtained by the SSS method over SS in the case of channels with closely spaced roots. The SSS method sounds to be a promising technique, yet it has a higher computational complexity that needs to be addressed in a future work.
\par
\textit{Notation}: The invertible column vector-matrix mappings are denoted by ${\rm{vec}} \{ .\} :\mathbb{C}{^{a\times b}} \to \mathbb{C}{^{ab \times 1}} $ and ${\rm{mat}}_{a,b}\{ .\} :\mathbb{C}{^{ab \times 1}} \to\mathbb{C}{^{a \times b}}$. $\left( {{\A} \otimes \B} \right)$ is the Kronecker product. $\A^T$ and $\A^H$ denote the transpose and Hermitian transpose, respectively.

\section{Problem Formulation}
\label{sec:format}

\subsection{Multi-channel model}
Multichannel framework is considered in this work. It is obtained either by oversampling the received signal or using an array of antennas or a combination of both \cite{tong1998multichannel}. To further develop the multi channel system model, consider the observed signal $y(t)$ from a linear modulation over a linear channel with additive noise given by	
\begin{equation}
y(t) = \sum\limits_k {h(t - k)} s(k) + e(t),\quad t = 0, \ldots ,N - 1
\label{model1}
\end{equation}
where $h(t)$ is the FIR channel impulse response, $s(k)$ are the transmitted symbols and $e(t)$ is the additive noise. If the received signal is oversampled or recorded with $m$ sensors, the signal model in (\ref{model1}) becomes $m$-variate and expressed as
\begin{equation}
\y(t) = \sum\limits_{i = 0}^{L-1} {\h(i)s(t - i) + \e(t)}
\label{model2}
\end{equation}
where $\y(t) = [y_1(t), \cdots ,y_m(t)]^T$, $\h(i) = {[h_1(i), \cdots ,h_m(i)]^T}$, $\e(t) = {[e_1(t), \cdots ,e_m(t)]^T}$. Define the system transfer function $\H(z) = \sum\nolimits_{k = 0}^{L-1} {\h(k){z^{ - k}}}$ with $(L-1)=deg(\H(z))$. Consider the noise to be additive independent white circular noise with $E[\e(k){\e^H}(i)] = \delta_{k,i}\sigma _e^2{\mathbf{I}_m}$. Assume reception of a window of $M$  samples, by stacking the data into a vector/matrix representation, we get:
\begin{equation}
\y_{M}(t) = {\cH}_M(\h)\s_{M+L-1}(t) + \e_M(t)
\label{model3}
\end{equation}
where $\y_M(t)={[\y^H(t), \cdots ,\y^H(t-M+1)]^H}$, $\s_{M+L-1}(t)= {[{s}(t), \cdots ,{s}(t-M -L+2 )]^T}$, $\e_M(t)$ is stacked in a similar way to as $\y_M(t)$, and ${\cH_M(\h)}$ is an $mM\times (M+L-1)$ block Toeplitz matrix defined as
\begin{align}
\label{eq_H}
\cH_M(\h) = \left[ {\begin{array}{*{20}{c}}
{\h(0)}& \cdots &{\h(L - 1)}& \cdots &0\\
 \vdots & \ddots & \ddots & \ddots & \vdots \\
0& \cdots &{\h(0)}& \cdots &{\h(L - 1)}
\end{array}} \right]
\end{align}
$\h$ is the desired parameter vector containing all channels taps, i.e. $\h=[\h(0)^T, \cdots, \h(L-1)^T]^T$. Using the observation data in (\ref{model3}),  our objective is to  estimate the different channels' impulse responses, i.e, recover $\h$ up to a possible scalar ambiguity. In the following subsection, we describe the subspace method, briefly.
\subsection{Subspace method revisited}
For consistency and reader's convenience, the SS method \cite{moulines1995subspace} which is also referred to as noise subspace method, shall be reviewed hereafter. The SS method implicitly exploits the Toeplitz structure of the filtering matrix $\cH_M(\h)$. Let $\v = [\v_1^H, \cdots ,\v_M^H]^H$, where $\v_i = [v_{(i-1)m+1}, \cdots ,v_{im}]^T,\; i = 1, \ldots ,M$, be in the orthogonal complement space of the range space of $\cH_M(\h)$ such that
\begin{align}
\label{eq_vH}
\v^H\cH_M(\h) =0
\end{align}
Using the block Toeplitz structure of ${\cH}_M(\h)$, the above linear equation can be written in terms of the channel parameter $\h$ as
\begin{align}
\label{eq_hV}
{\h^H}\left[ {\begin{array}{*{20}{c}}
{{\v_1}}& \cdots &{{\v_M}}&{0}&{0}\\
{0}& \ddots &{}& \ddots &{0}\\
{0}&{0}&{{\v_1}}& \cdots &{{\v_M}}
\end{array}} \right] = {\h^H}\cV = 0
\end{align}
The former equation can be used to estimate the channel vector $\h$ provided that \eqref{eq_hV} has a unique solution. Moulines et al. \cite{moulines1995subspace} proposed the SS method which is based on the following theorem:
\begin{theorem}
\label{theorem1}
Assume that the components of $\H(z)$ have no common zeros, and $M \ge L $. Let $\left\{{\v_i}\right\}_{i=1}^{d}$ be a basis of the orthogonal complement of the column space of $\cH_M(\h)$, then for any $\H'(z)$ with $deg(\H'(z))=L-1$ we have
\begin{align}
\cV_i^H\h'=0 , {\rm for}~~ i=1, \cdots, d \iff \H'(z)=\alpha\H(z)
\end{align}
where $\alpha$ is some scalar factor.
\end{theorem}
One of the encountered ways to estimate the orthogonal complement of $\cH_M(\h)$, i.e. noise subspace, is the signal-noise subspace decomposition. From the multi-channel model and noise properties, the received signal covariance matrix ${\R}_y=E[\y_M(t) \y_M^H(t)]$ is given as 
\begin{align}
{\R}_y = \cH_M(\h) {\bf R}_s \cH_M^H(\h) + \sigma _e^2 \I
\end{align}
The singular value decomposition of ${\R}_y$ has the form
\begin{align}
{\R_y} = {\V_s}\rm{diag}\left( {\lambda _1^2, \cdots ,\lambda _{M + L - 1}^2} \right)\V_s^H + \sigma _e^2{\V_e}\V_e^H
\end{align}
where $\lambda_i^2$, $i=1, \cdots, M+L-1$ are the principal eigenvalues of the covariance matrix ${\bf R}_y$. Also, the columns of $\V_s$ and $\V_e$ span the so-called signal and noise subspaces (orthogonal complement), respectively. 
After having the basis of the noise subspace, the channel identification can be performed based on the following quadratic optimization criterion:
\begin{align}
\hat \h = \arg \; \min \|\V_e^H{\cH_M(\h)}\|^2= \arg \;\min {\h^H}\left[ {\sum\limits_i {{\cV_i}\cV_i^H} } \right]\h
\label{eq:optSS}
\end{align}

In brief, the SS method achieves the channel estimation by exploiting the subspace information (i.e. ideally, (${\rm Range}{\cH}_M(\h))= {\rm Range}(\V_s) ~\bot ~{\rm Range}(\V_e)$) as well as the block Sylvester (block-Toeplitz) structure of the channel matrix. More precisely, it enforces the latter matrix structure through the use of relations \eqref{eq_vH} and \eqref{eq_hV} and minimizes the subspace orthogonality error in \eqref{eq:optSS}. In the sequel, we propose a dual approach which enforces the subspace information (i.e. ${\rm Range}({\cH}_M(\h))= {\rm Range}(\V_s)$ where $\V_s$ refers to the principal subspace of the sample covariance matrix) while minimizing a cost function representing  the deviation of ${\cH}_M(\h)$ from the Sylvester structure as indicated in Table \ref{Tab:Duality Table}.

\begin{table}[!t]
\renewcommand{\arraystretch}{1.3}
\caption{Duality Table}
\label{Tab:Duality Table}
\centering
\begin{tabular}{|l||c|c|}
\hline
\textbf{Method} &  \textbf{Toeplitz Structure} & \textbf{Orthogonality}\\
\hline \hline
\textbf{SS }& forced & minimized \\
\hline
\textbf{SSS }& minimized & forced \\
\hline
\end{tabular}
\end{table}

\section{Structure-Based SS method (SSS)}
\label{sec:pagestyle}

In the proposed subspace method, one searches for the system matrix ${\cH}_M$ in the form $\hat{\cH}_M= \V_s\Q$ so that the orthogonality criterion in \eqref{eq:optSS} is set equal to zero, i.e. $\|\V_e^H\hat{\cH}_M\|^2=0$ while $\Q$ is chosen in such a way the resulting matrix is close to the desired block Toepliz structure. This is done by minimizing w.r.t. $\Q$ the following structure-based cost function (informal Matlab
notions are used):

\begin{align}
\label{eq_C}
\begin{array}{l}
\cJ = {\cJ_1} + {\cJ_2} + {\cJ_3}\\
 = \;{\left| {\sum\limits_{j = 1}^{K - 1} {\sum\limits_{i = 1}^{m(M - 1)} {\hat w(i,j) - \hat w(i + m,j + 1)} } } \right|^2}\\
 + {\left| {\sum\limits_{j = L + 1}^{K} {\hat w(1:m,j)} } \right|^2} + {\left| {\sum\limits_{i = m + 1}^{mM} {\hat w(i,1)} } \right|^2}
\end{array}
\end{align}
where $K=M+L-1$ and $\hat{\W}$ refers to $\hat{\cH}_M$. The cost function in \eqref{eq_C} is inspired and matched to the Toeplitz structure introduced in \eqref{eq_H}. It is a composite of three parts; $\cJ_1$ seeks to force Toeplitz structure on the possibly non-zero entries, while $\cJ_2$ and $\cJ_3$ account for the zero entries in the first $ m $ rows and first column, respectively.
\par
Starting with $\cJ_1$, one can express it in a more compact way as follows:
\begin{align}
{{\cJ_1} = {{\| {{\I_L}\hat \W{\I_R} - {\J_L}\hat \W{\J_R}} \|}^2}}
\end{align} 
where:\\
$\I_L$ is the $(mM) \times (mM)$ left identity square matrix with setting the last $m$ diagonal entries to zeros. \\
$\I_R$ is the $K \times K$ right identity square matrix with setting the last diagonal entry to zero.  \\
$\J_L $ is a $(mM) \times (mM)$ square translation matrix with ones on the sub-diagonal and zeros elsewhere, i.e., ${\left[ {{J_L}} \right]_{i,j}} = {\delta _{i + {m},j}}$. \\
$\J_R $ is a $K \times K$ square translation matrix with ones on the super-diagonal and zeros elsewhere, i.e., ${\left[ {{J_R}} \right]_{i,j}} = {\delta _{i,j+1}}$.
\par
Now, using the Kronecker product property ${\rm{vec}}(\A\G\B) = \left( {{\B^T} \otimes \A} \right){\rm{vec}}(\G) = \left( {{\B^T} \otimes \A} \right)\g$, one can write $\cJ_1$ as follows:
\begin{align}
\label{eq_C1}
\begin{array}{l}
{\cJ_1} = {\| {\left( {{\I_R} \otimes {\I_L} - \J_R^T \otimes {\J_L}} \right){\rm{vec}}(\hat \W)} \|^2}\\
\quad  = {\| {\left( {{\I_R} \otimes {\I_L} - \J_R^T \otimes {\J_L}} \right)\left( {\I \otimes \V_s} \right)\q} \|^2} = {\| {{\K_1}\q} \|^2}
\end{array}
\end{align}
where $\q={\rm{vec}}(\Q)$.
In a similar way, $\cJ_2$ can be expressed as
\begin{align}
\label{eq_C2}
\begin{array}{l}
{\cJ_2} = {\| {\hat \W\left( {1:{m},L + 1:end} \right)} \|^2} = \|{\V_{s,row}}\Q{\I_{row}}\|^2\\
\quad \quad  = {\| {\left( {{\I_{row}} \otimes {\V_{s,row}}} \right)\q} \|^2} = {\| {{\K_2}\q} \|^2}
\end{array}
\end{align}
where
$\V_{s,row}$ is the sub-matrix of $\V_s$ given by its first $m$ rows, and $\I_{row}$ is the $K\times K$ square identity matrix with setting the first $L$ diagonal entries to zero. Finally, $\cJ_3$ can also be set up as
\begin{align}
\label{eq_C3}
\begin{array}{l}
{\cJ_3} = {\| {\hat \W\left( {m + 1:mM,1} \right)} \|^2} = \|{\V_{s,col}}\Q{\I_{col}}\|^2\\
\quad \quad  = {\| {\left( {{\I_{col}} \otimes {\V_{s,col}}} \right)\q} \|^2} = {\| {{\K_3}\q} \|^2}
\end{array}
\end{align}
where
$\V_{s,col}$ is the sub-matrix of $\V_s$ given by its last $m(M-1)$ rows, and $\I_{col}$ is the $K\times K$ square diagonal matrix with one at the first diagonal entry and zeros elsewhere.
\par
As a result of \eqref{eq_C1}, \eqref{eq_C2} and \eqref{eq_C3} the optimization problem in \eqref{eq_C} is reduced to the minimization of the following quadratic equation
\begin{align}
\label{eq_min}
{\mathop {\min }\limits_\q \quad {\q^H}\K^H\K\q}
\end{align} 
where $\K = {\left[ {\begin{array}{c|c|c}
{\K_1^T}&{\K_2^T}&{\K_3^T}
\end{array}} \right]^T}$.
\par
 The optimal solution $\q$ of \eqref{eq_min}, under unit norm constraint of $\q$, is the least eigenvector that corresponds to the smallest eigenvalue of $\K^H\K$. The square matrix $\Q$ can be constructed by reshaping the obtained solution $\q$ from a vector into the matrix format, such that $\Q={\rm{mat}}_{K,K}\{ \q\}$. 
 Once matrix $\Q$ is obtained, the channel taps are estimated by averaging over the non-zero diagonal blocks of matrix $\V_s\Q$.

\section{Discussion}
In this section, we provide some insightful comments in order to highlight the advantages and drawbacks of the proposed subspace method.
\begin{itemize}

\item As explained earlier the proposed  approach consists of neglecting the subspace error (i.e. considering Range($\V_s$) as perfect in the sense one searches for the desired solution within that subspace) while minimizing the system matrix  (Toeplitz) structure error. The motivation behind this choice resides in the fact that the subspace error at the first order is null as shown in \cite{xu2002perturbation} and hence it can be neglected at the first order in favor of more flexibility for searching the appropriate channel matrix. This explains the observed gain of the SSS over SS method in certain difficult scenarii including the case of closely spaced channels roots. 

\item In the favorable cases where the channel matrix is well conditioned, the two subspace methods lead to similar performance as illustrated next by the simulation example of Fig. \ref{fig_MSE_VS_SNR_well_1}.

\item For the SS method to apply one needs that the noise subspace vectors generate a minimal polynomial basis of the rational subspace orthogonal to ${\rm Range}(\H(z))$ (see \cite{moulines1995subspace} for more details) and so the condition $M \geq  L$ is considered to guarantee such requirement to hold. As the SSS does not explicitly rely on the orthogonality relation in \eqref{eq:optSS}, the latter condition might be relaxed as illustrated by the simulation example of Fig. \ref{fig_MSE_VS_SNR_wind_can35_1}.

\item The proposed subspace method has a higher numerical cost as compared to the SS method.  However, the cost might be reduced by taken into account the  Kronecker products involved in building matrix $\K$. This issue is still under investigation together with an asymptotic  statistical performance analysis of SSS.

\item In the case $M \geq L$, the solution of \eqref{eq_min} can be shown to be unique (up to a constant) thanks to the identifiability result of Theorem \ref{theorem1}. Indeed, if $\q'$ is another solution zeroing criterion \eqref{eq_C}, then the FIR filter associated to matrix $\cH'=\V_s\Q'$ satisfies all conditions of Theorem \ref{theorem1}, which leads to $\V_s\Q'= \alpha \V_s\Q $ or equivalently $\Q'= \alpha \Q$.

\end{itemize}

\section{SIMULATION RESULTS}
\label{sec:simulation}
In this section, the devised SSS method will be compared to the standard SS method as a benchmark. Three different experiments will be examined to illustrate the behavior of SSS in different contexts.


Two FIR channels are considered, each has a second order impulse response given by \cite{qiu1996performance}:
\begin{align*}
\begin{array}{l}
{{\bf{h}}_1} = {\left[ {\begin{array}{*{20}{c}}
1&{ - 2\cos (\theta )}&1
\end{array}} \right]^T},\\
{{\bf{h}}_2} = {\left[ {\begin{array}{*{20}{c}}
1&{ - 2\cos (\theta  + \delta )}&1
\end{array}} \right]^T}
\end{array}
\end{align*}
where $\theta$ is the absolute phase value of ${\bf{h}}_1$'s zeros and $\delta$ indicates the angular distance between the zeros of the two channels on the unit circle. Small $\delta$ results into an ill-conditioned system. In all simulations, the excitation signal is a 4-QAM, each channel receives $N=100$ samples, and the noise is additive white Gaussian. Note that the SNR is defined as
\begin{align*}
SNR(dB) = 10\log_{10} E\frac{{{{\left\| {\cH_N\s_{N+L-1}} \right\|}^2}}}{{mN\sigma _v^2}}
\end{align*}
The performance measure is the mean-square-error (MSE), given as
\begin{align*}
MSE(dB) = 20lo{g_{10}}\left( {\frac{1}{{||\h||}}\sqrt {\frac{1}{{{N_{mc}}}}\sum\limits_{i = 1}^{{N_{mc}}} {||{\hat\h_i} - \h|{|^2}} } } \right)
\end{align*}
where $N_{mc}=100$ refers to the number of Monte-Carlo runs and $\hat{\h}_i$ is the channel vector estimate at the $i$-th run.
\par
\begin{figure}[!t]
\centering
\includegraphics[width= 7cm, height= 5cm]{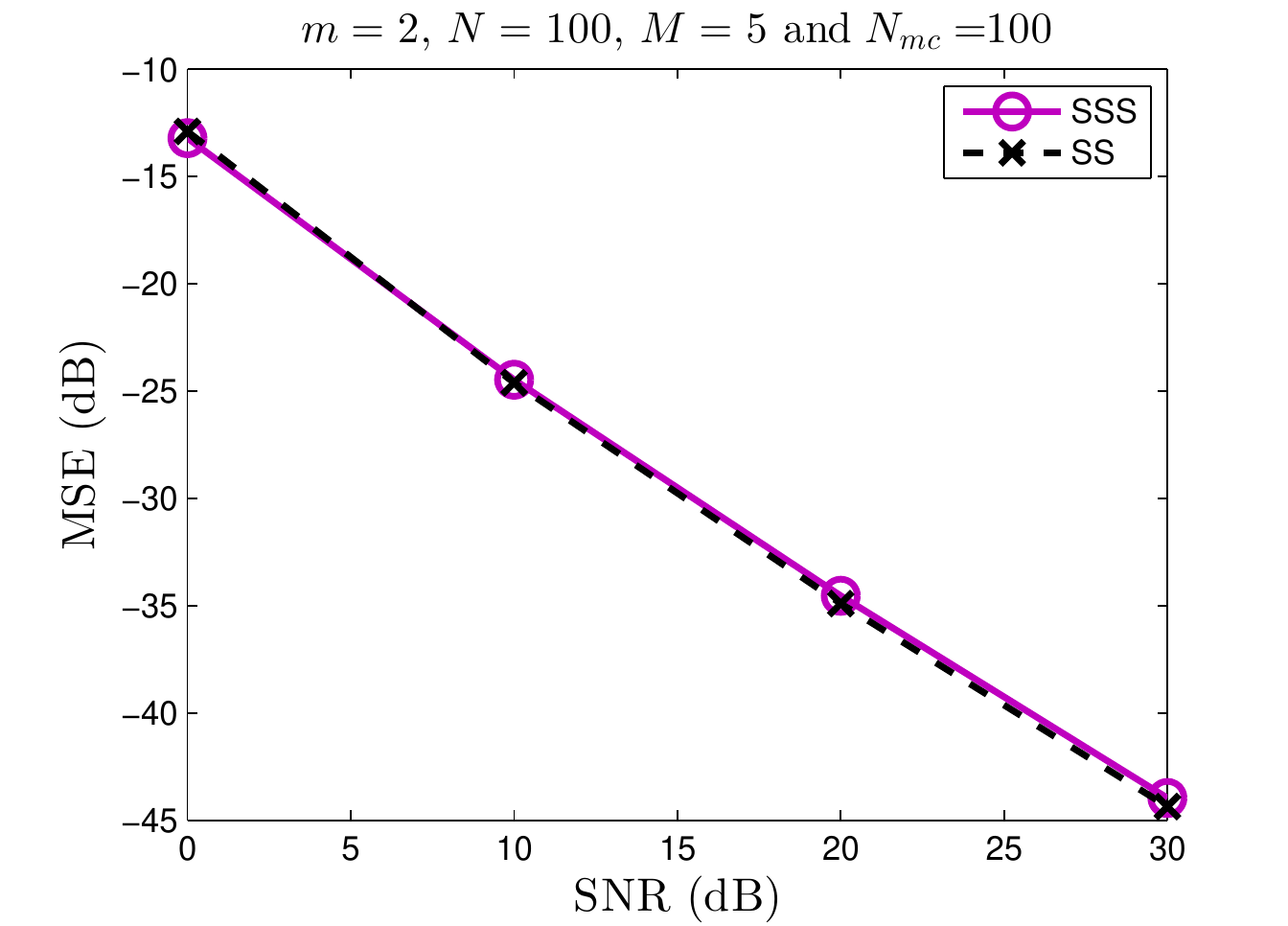}
\caption{Well-conditioned channels, $\theta=\pi/10,\delta=\pi$.}
\label{fig_MSE_VS_SNR_well_1}
\end{figure}
\begin{figure}[!t]
\centering
\includegraphics[width= 7cm, height= 5cm]{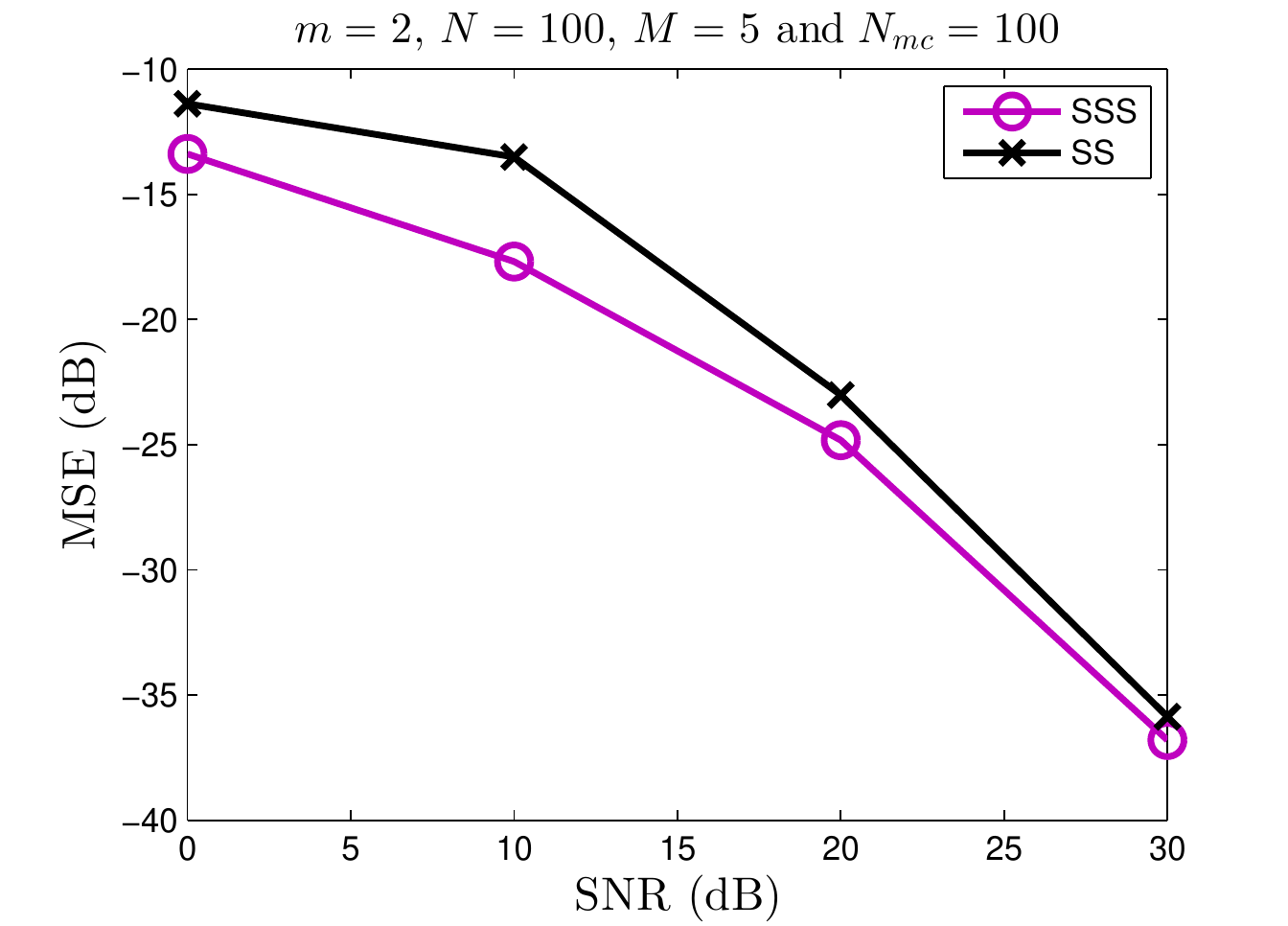}
\caption{Ill-conditioned system, $\theta=\pi/10,\delta=\pi/10$.}
\label{fig_MSE_VS_SNR_ill_1}
\end{figure}
\begin{figure}[!t]
\centering
\includegraphics[width= 7cm, height= 5cm]{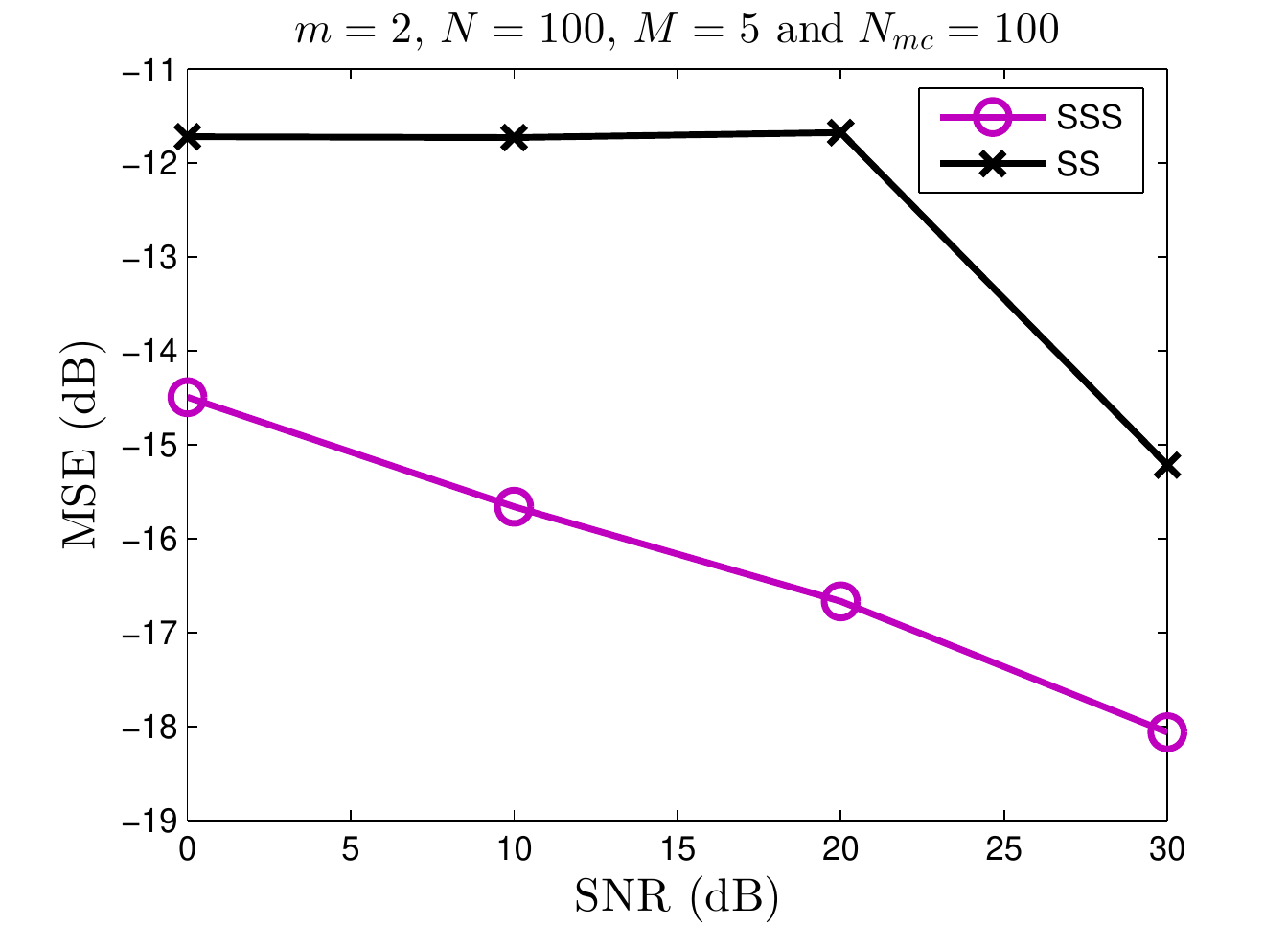}
\caption{Severely ill-conditioned system, $\theta=\pi/10,\delta=\pi/50$.}
\label{fig_MSE_VS_SNR_ill_2}
\end{figure}
\begin{figure}[!t]
\centering
\includegraphics[width= 7cm, height= 5cm]{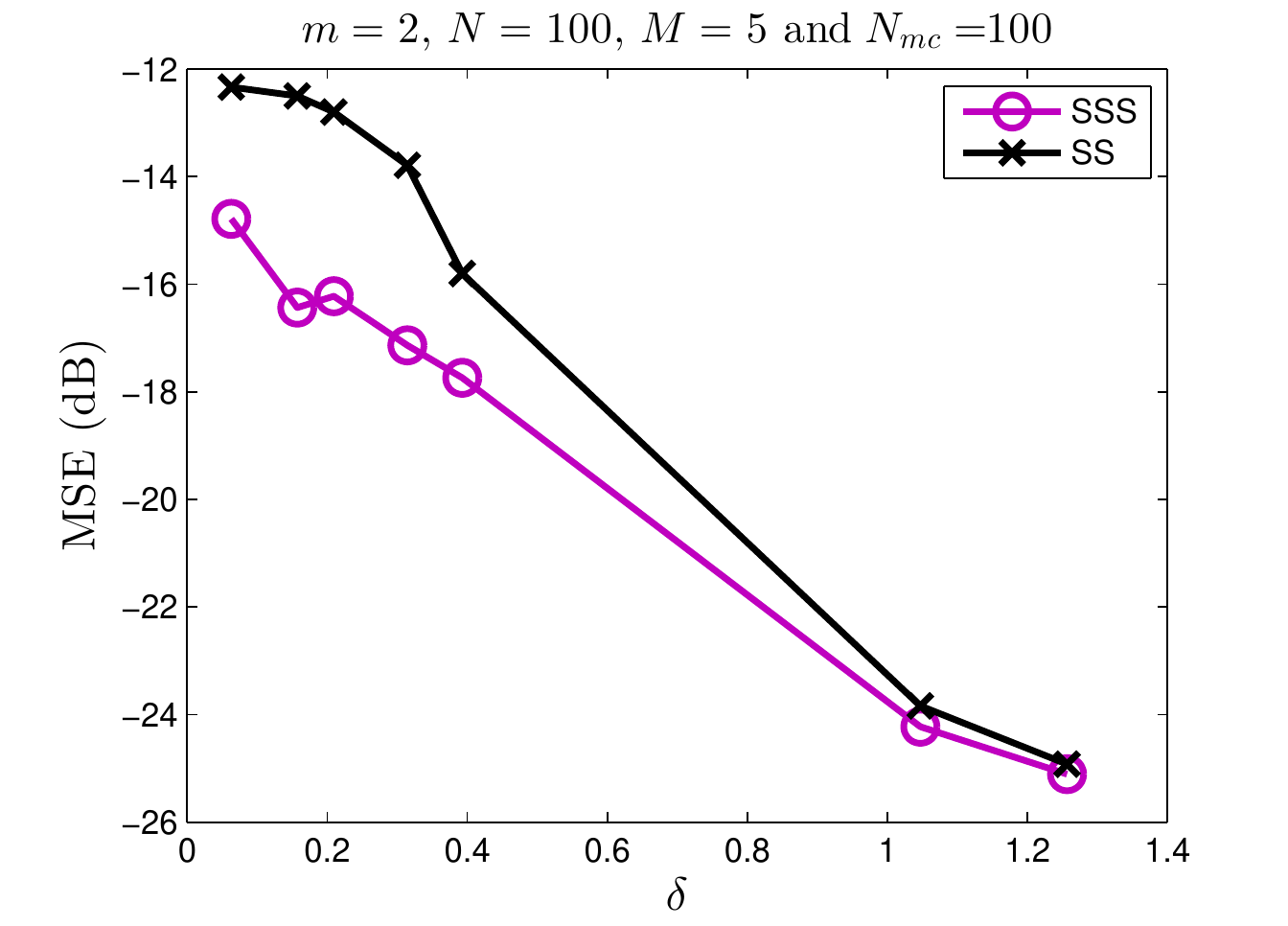}
\caption{MSE versus $\delta$, $SNR=10 \; dB$}
\label{fig_MSE_VS_delta2}
\end{figure}

In the first experiment given by Fig. \ref{fig_MSE_VS_SNR_well_1}, we show that for a well-conditioned system ($\delta=\pi$), both methods have a comparable performance. In the second one, we consider  ill-conditioned systems (i.e. poor channel diversity). In that case, the devised SSS method outperforms the SS method at a moderate ill-conditioned system ($\delta=\pi/10$), and its performance gain becomes more obvious at severely ill-conditioned case ($\delta=\pi/50$) as shown in Fig.'s \ref{fig_MSE_VS_SNR_ill_1} and \ref{fig_MSE_VS_SNR_ill_2}, respectively. 
When the system is ill-conditioned, the difference becomes more pronounced at low and moderate SNR values. Also, at the severe ill-conditioned case, the SS methods becomes unresponsive to the changes in the signal's SNR, as revealed in Fig. \ref{fig_MSE_VS_SNR_ill_2}, since the effect of ill-conditioning becomes prominent at low SNR. Figure \ref{fig_MSE_VS_delta2} depicts the consequence of varying $\delta$ on the MSE for SNR=10dB.

\begin{figure}[!t]
\centering
\includegraphics[width= 7cm, height= 5cm]{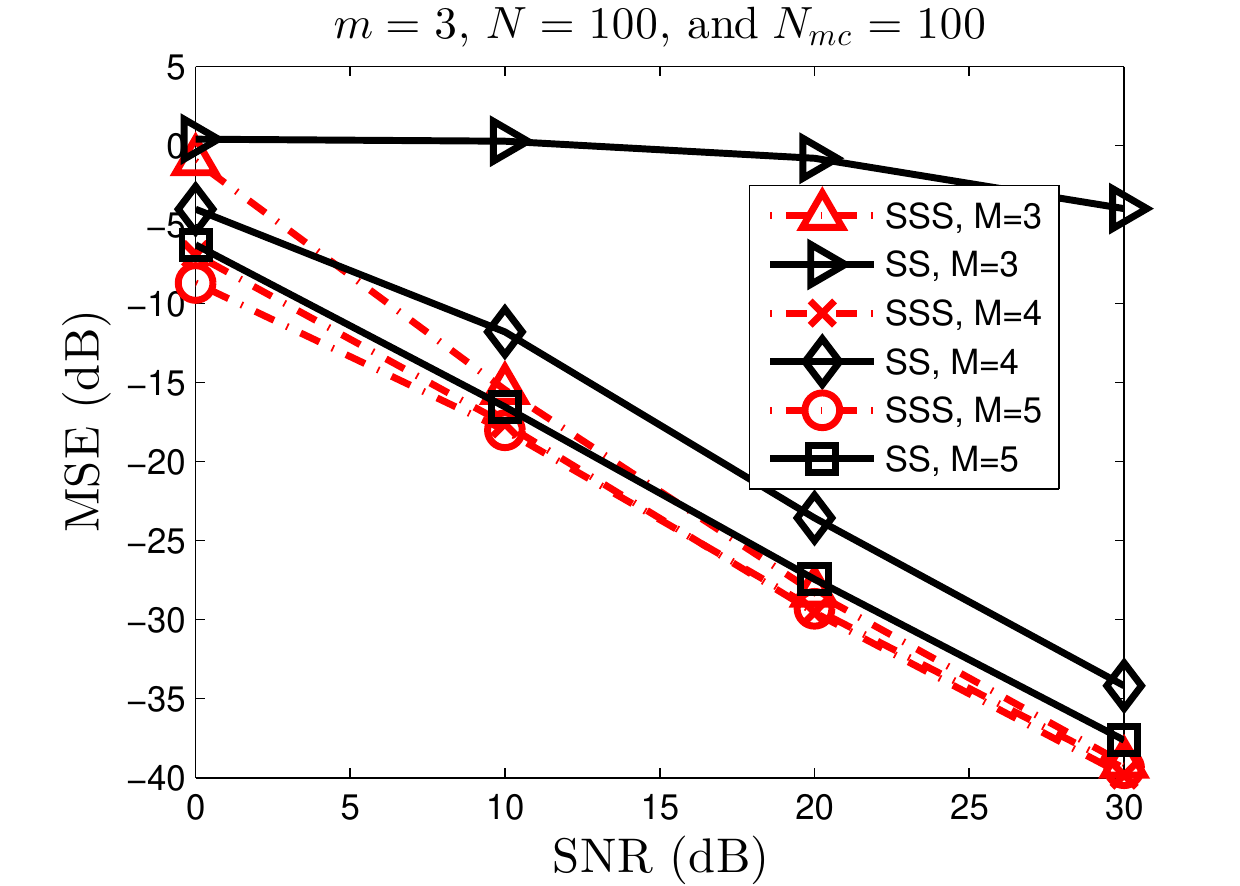}
\caption{MSE versus SNR for different window size $M$.}
\label{fig_MSE_VS_SNR_wind_can35_1}
\end{figure}

In the last experiment, the number of channels is $m=3$ and the number of taps in each channel is $L=5$, the transfer function of the channels are given in \cite{abed1997subspace}. In this experiment, we are primarily interested to look at the impact of the processing window length on the estimation performance. As can be seen from the results reported in Fig. \ref{fig_MSE_VS_SNR_wind_can35_1}, the performance of the SS method gets worse and degrades when the processing window length $M$ becomes less than the number of the channels' taps $L$, while our proposed SSS is weakly affected by the window length condition, i.e. $M \ge L$. This allows us to reduce the dimension of the channel matrix ${\cH}_M$ with smaller window size values, especially for large dimensional systems where  $m \gg 1$.

\section{CONCLUSION}
\label{sec:conclusion}
In this paper, we proposed a dual approach to the standard subspace method, whereby the channel matrix is forced to belong to the principal subspace of the data covariance matrix estimate while its deviation from Toeplitz structure is minimized. By doing so, we show that the channel estimation is significantly improved in the difficult context of weak channels diversity (i.e. channels with closely spaced roots). Interestingly, the principle of the proposed approach can be applied for estimation problems with other matrix structures where subspace method can be used.

\ifCLASSOPTIONcaptionsoff
  \newpage
\fi

\bibliographystyle{IEEEtran}
\bibliography{SPL}

%




\end{document}